\documentclass[aps,preprint,nofootinbib]{revtex4}%
\usepackage{amssymb}
\usepackage{amsmath}
\usepackage{epsfig}
\usepackage{amsfonts}
\usepackage{graphicx}
\usepackage{hyperref}%
\begin{document}
\title{Patterns of High energy Massive String Scatterings in the Regge Regime}
\author{Sheng-Lan Ko}
\email{slko.py96g@g2.nctu.edu.tw}
\affiliation{Department of Electrophysics, National Chiao-Tung University, Hsinchu, Taiwan,
R.O.C. }
\author{Jen-Chi Lee}
\email{jcclee@cc.nctu.edu.tw}
\affiliation{Department of Electrophysics, National Chiao-Tung University and Physics
Division, National Center for Theoretical Sciences, Hsinchu, Taiwan, R.O.C.}
\author{Yi Yang}
\email{yiyang@mail.nctu.edu.tw}
\affiliation{Department of Electrophysics, National Chiao-Tung University and Physics
Division, National Center for Theoretical Sciences, Hsinchu, Taiwan, R.O.C.}
\date{\today }

\begin{abstract}
We calculate high energy massive string scattering amplitudes of open bosonic
string in the Regge regime (RR). We found that the number of high energy
amplitudes for each fixed mass level in the RR is much more numerous than that
of Gross regime (GR) calculated previously. Moreover, we discover that the
leading order amplitudes in the RR can be expressed in terms of the Kummer
function of the second kind. In particular, based on a summation algorithm for
Stirling number identities developed recently, we discover that the ratios
calculated previously among scattering amplitudes in the GR can be extracted
from this Kummer function in the RR. We conjecture and give evidences that the
existence of these GR ratios in the RR persists to subleading orders in the
Regge expansion of all string scattering amplitudes. Finally, we demonstrate
the universal power-law behavior for all massive string scattering amplitudes
in the RR.

\end{abstract}
\maketitle
\tableofcontents
%

\setcounter{equation}{0}
\renewcommand{\theequation}{\arabic{section}.\arabic{equation}}%

\section{Introduction}

There are two fundamental regimes of high energy string scattering amplitudes.
These are the fixed angle regime or Gross regime (GR), and the fixed momentum
transfer regime or Regge regime (RR). These two regimes represent two
different high energy perturbation expansions of the scattering amplitudes,
and contain complementary information of the theory. The UV behavior of high
energy string scatterings in the GR is well known to be very soft exponential
fall-off, while that of RR is hard power-law. The high energy string
scattering amplitudes in the GR \cite{GM, Gross, GrossManes} were recently
intensively reinvestigated for massive string states at arbitrary mass levels
\cite{ChanLee1,ChanLee2, CHL,CHLTY,PRL,susy,Closed,Decay,Compact}. See also
the developments in \cite{West1,West2,Moore}. An infinite number of linear
relations, or stringy symmetries, among string scattering amplitudes of
different string states were obtained. Moreover, these linear relations can be
solved for each fixed mass level, and ratios $T^{(N,2m,q)}/T^{(N,0,0)}$ among
the amplitudes can be obtained. An important new ingredient of these
calculations is the decoupling of zero-norm states (ZNS) \cite{ZNS1,ZNS3,ZNS2}
in the old covariant first quantized (OCFQ) string spectrum. It is interesting
to note that the calculation in \cite{GM, Gross, GrossManes} is valid only for
four-tachyon amplitude, but not for all other amplitudes of excited string
states. This was pointed out and the calculation was corrected by two
independent groups \cite{CHL,West2} with two different approaches. Since there
does not exist any algebraic structure (or group structure) of this high
energy 26D spacetime symmetry, mathematically the meaning of these infinite
number of ratios remains mysterious.

Another fundamental regime of high energy string scattering amplitudes is the
RR \cite{RR1,RR2,RR3,RR4,RR5,RR6}. See also \cite{OA,DL,KP}. Since the
decoupling of ZNS applies to all kinematic regimes, one expects some
implication of this decoupling in the RR. Moreover, it is conceivable that
there exists some link between the patterns of the high energy scattering
amplitudes of GR and RR. With this in mind, in this paper, we give a detail
calculation of high energy string scattering amplitudes in the RR. We will
find that the number of high energy scattering amplitudes for each fixed mass
level in the RR is much more numerous than that of GR calculated previously.
On the other hand, it seems that both the saddle-point method and the method
of decoupling of high energy ZNS adopted in the calculation of GR do not apply
to the case of RR. However the calculation is still manageable, and the
general formula for the high energy scattering amplitudes for each fixed mass
level in the RR can be written down explicitly.

In contrast to the case of scatterings in the GR, we will see that there is no
linear relation among scatterings in the RR. Moreover, we discover that the
leading order amplitudes at each fixed mass level in the RR can be expressed
in terms of the Kummer function of the second kind. More surprisingly, for
those leading order high energy amplitudes $A^{(N,2m,q)}$ in the RR with the
same type of $(N,2m,q)$ as those of GR, we can extract from them the ratios
$T^{(N,2m,q)}/T^{(N,0,0)}$ in the GR by using this Kummer function.
Mathematically, the proof of this result turns out to be highly nontrivial and
is based on a summation algorithm for Stirling number identity derived by
Mkauers in 2007 \cite{MK}. It is very interesting to see that the identity in
Eq.(\ref{ID}) suggested by string theory calculation can be rigorously proved
by a totally different but sophisticated mathematical method. The derivation
of these physical ratios from Kummer function through Stirling number
identities seems to suggest another interpretation of these infinite number of
ratios mathematically. We then proceed to calculate Regge string scattering
amplitudes to subleading orders. We conjecture and give evidences that these
ratios persist to all orders in the Regge expansion of high energy string
scattering amplitudes for the even mass level with $(N-1)=\frac{M_{2}^{2}}{2}%
$= even. For the odd mass levels with $(N-1)=\frac{M_{2}^{2}}{2}$= odd, the
existence of the GR ratios shows up only in the first $[N/2]+1$ terms in the
Regge expansion of the amplitudes. At last, as an application of our results,
we show that the well known $s^{\alpha(t)}$ power-law behavior of the four
tachyon string scattering amplitude in the RR can be extended to all high
energy massive string scattering amplitudes.

This paper is organized as following. In section II, after a brief review of
high energy string scatterings in the GR, we first calculate all leading high
energy scattering amplitudes for the mass level $M^{2}=4$ in the RR. We
compare the two sets of amplitudes and discover a link between the two. The
calculation is then generalized to general mass level in the RR in section
III. We show that the leading order amplitudes can be expressed in terms of
the Kummer function of the second kind. In section IV, based on a summation
algorithm for Stirling number identity, we show that the ratios among
scattering amplitudes in the GR can be extracted from Kummer function derived
in section III. In section V, we give evidences that the existence of these
ratios in the RR persists to subleading orders in the Regge expansion of all
high energy string scattering amplitudes. In section VI, we demonstrate the
universal power-law behavior for all massive string scattering amplitudes in
the RR. Finally, an appendix is devoted to the kinematics used in the text.%

\setcounter{equation}{0}
\renewcommand{\theequation}{\arabic{section}.\arabic{equation}}%

\section{Regge Scattering for $M_{2}^{2}=4$}

We begin with a brief review of high energy string scatterings in the GR. That
is in the kinematic regime%
\begin{equation}
s,-t\rightarrow\infty,t/s\approx-\sin^{2}\frac{\theta}{2}=\text{fixed (but
}\theta\neq0\text{)}%
\end{equation}
where $s,t$ and $u$ are the Mandelstam variables and $\theta$ is the CM
scattering angle. It was shown \cite{CHLTY,PRL} that for the 26D open bosonic
string the only states that will survive the high-energy limit at mass level
$M_{2}^{2}=2(N-1)$ are of the form%
\begin{equation}
\left\vert N,2m,q\right\rangle \equiv(\alpha_{-1}^{T})^{N-2m-2q}(\alpha
_{-1}^{L})^{2m}(\alpha_{-2}^{L})^{q}|0\rangle, \label{relevant states}%
\end{equation}
where the polarizations of the 2nd particle with momentum $k_{2}$ on the
scattering plane were defined to be $e^{P}=\frac{1}{M_{2}}(E_{2}%
,\mathrm{k}_{2},0)=\frac{k_{2}}{M_{2}}$ as the momentum polarization,
$e^{L}=\frac{1}{M_{2}}(\mathrm{k}_{2},E_{2},0)$ the longitudinal polarization
and $e^{T}=(0,0,1)$ the transverse polarization. Note that $e^{P}$ approaches
to $e^{L}$ in the GR, and the scattering plane is defined by the spatial
components of $e^{L}$ and $e^{T}$. Polarizations perpendicular to the
scattering plane are ignored because they are kinematically suppressed for
four point scatterings in the high-energy limit. One can use the saddle-point
method to calculate the high energy scattering amplitudes. For simplicity, we
choose $k_{1}$, $k_{3}$ and $k_{4}$ to be tachyons and the final result of the
ratios of high energy, fixed angle string scattering amplitude are
\cite{CHLTY,PRL}%
\begin{equation}
\frac{T^{(N,2m,q)}}{T^{(N,0,0)}}=\left(  -\frac{1}{M_{2}}\right)
^{2m+q}\left(  \frac{1}{2}\right)  ^{m+q}(2m-1)!!. \label{ratios}%
\end{equation}
The ratios in Eq.(\ref{ratios}) can also be obtained by using the decoupling
of two types of ZNS in the spectrum%
\begin{equation}
\text{Type I}:L_{-1}\left\vert x\right\rangle ,\text{ where }L_{1}\left\vert
x\right\rangle =L_{2}\left\vert x\right\rangle =0,\text{ }L_{0}\left\vert
x\right\rangle =0;
\end{equation}%
\begin{equation}
\text{Type II}:(L_{-2}+\frac{3}{2}L_{-1}^{2})\left\vert \widetilde
{x}\right\rangle ,\text{ where }L_{1}\left\vert \widetilde{x}\right\rangle
=L_{2}\left\vert \widetilde{x}\right\rangle =0,\text{ }(L_{0}+1)\left\vert
\widetilde{x}\right\rangle =0.
\end{equation}
As examples, for $M_{2}^{2}=4,6$, we get \cite{ChanLee1,ChanLee2}
\begin{equation}
T_{TTT}:T_{LLT}:T_{(LT)}:T_{[LT]}=8:1:-1:-1, \label{CL}%
\end{equation}%
\begin{equation}%
\begin{array}
[c]{ccccccccccccccccc}%
T_{TTTT} & : & T_{TTLL} & : & T_{LLLL} & : & T_{TTL} & : & T_{LLL} & : &
\tilde{T}_{LT,T} & : & \tilde{T}_{LP,P} & : & T_{LL} & : & \tilde{T}_{LL}\\
16 & : & \frac{4}{3} & : & \frac{1}{3} & : & -\frac{4\sqrt{6}}{9} & : &
-\frac{\sqrt{6}}{9} & : & -\frac{2\sqrt{6}}{3} & : & 0 & : & \frac{2}{3} & : &
0
\end{array}
.. \label{CL2}%
\end{equation}
\qquad\qquad

We now turn to the discussion on high energy string scatterings in the RR.
That is in the kinematic regime%
\begin{equation}
s\rightarrow\infty,\sqrt{-t}=\text{fixed (but }\sqrt{-t}\neq\infty).
\end{equation}
As in the case of GR, we only need to consider the polarizations on the
scattering plane, which is defined in Appendix A. Appendix A also includs the
kinematic set up and some formulas we need in our calculation. However,
instead of using $(E,\theta)$ as the two independent kinematic variables in
the GR, we choose to use $(s,t)$ in the RR. One of the reason has been, in the
RR, $t\sim E\theta$ is fixed, and it is more convenient to use $(s,t)$ rather
than $(E,\theta).$ In the RR, to the lowest order, equations (\ref{A13}) to
(\ref{A18}) reduce to%
\begin{subequations}
\begin{align}
e^{P}\cdot k_{1}  &  =-\frac{1}{M_{2}}\left(  \sqrt{p^{2}+M_{1}^{2}}%
\sqrt{p^{2}+M_{2}^{2}}+p^{2}\right)  \simeq-\frac{s}{2M_{2}},\\
e^{L}\cdot k_{1}  &  =-\frac{p}{M_{2}}\left(  \sqrt{p^{2}+M_{1}^{2}}%
+\sqrt{p^{2}+M_{2}^{2}}\right)  \simeq-\frac{s}{2M_{2}},\\
e^{T}\cdot k_{1}  &  =0
\end{align}
and%
\end{subequations}
\begin{subequations}
\begin{align}
e^{P}\cdot k_{3}  &  =\frac{1}{M_{2}}\left(  \sqrt{q^{2}+M_{3}^{2}}\sqrt
{p^{2}+M_{2}^{2}}-pq\cos\theta\right)  \simeq-\frac{\tilde{t}}{2M_{2}}%
\equiv-\frac{t-M_{2}^{2}-M_{3}^{2}}{2M_{2}},\\
e^{L}\cdot k_{3}  &  =\frac{1}{M_{2}}\left(  p\sqrt{q^{2}+M_{3}^{2}}%
-q\sqrt{p^{2}+M_{2}^{2}}\cos\theta\right)  \simeq-\frac{\tilde{t}^{\prime}%
}{2M_{2}}\equiv-\frac{t+M_{2}^{2}-M_{3}^{2}}{2M_{2}},\\
e^{T}\cdot k_{3}  &  =-q\sin\phi\simeq-\sqrt{-{t}}.
\end{align}
Note that $e^{P}$ \textit{does not} approach to $e^{L}$ in the RR. This is
very different from the case of GR. In the following discussion, we will
calculate the amplitudes for the longitudinal polarization $e^{L}.$ For the
$e^{P}$ amplitudes, the results can be trivially modified. There is another
important difference between the high energy scattering amplitudes in the RR
and in the GR. We will find that the number of high energy scattering
amplitudes for each fixed mass level in the RR is much more numerous than that
of GR calculated previously. On the other hand, it seems that both the
saddle-point method and the method of decoupling of high energy ZNS adopted in
the calculation of GR do not apply to the case of RR. In this section, we will
explicitly calculate the string scattering amplitudes on the scattering plane
$\left(  e^{L},e^{T}\right)  $ for the mass level $M_{2}^{2}=4$. In the mass
level $M_{2}^{2}=4$ $\left(  M_{1}^{2}=M_{3}^{2}=M_{4}^{2}=-2\right)  $, it
turns out that there are eight high energy amplitudes in the RR%
\end{subequations}
\begin{align}
&  \alpha_{-1}^{T}\alpha_{-1}^{T}\alpha_{-1}^{T}|0\rangle,\alpha_{-1}%
^{L}\alpha_{-1}^{T}\alpha_{-1}^{T}|0\rangle,\alpha_{-1}^{L}\alpha_{-1}%
^{L}\alpha_{-1}^{T}|0\rangle,\alpha_{-1}^{L}\alpha_{-1}^{L}\alpha_{-1}%
^{L}|0\rangle,\nonumber\\
&  \alpha_{-1}^{T}\alpha_{-2}^{T}|0\rangle,\alpha_{-1}^{T}\alpha_{-2}%
^{L}|0\rangle,\alpha_{-1}^{L}\alpha_{-2}^{T}|0\rangle,\alpha_{-1}^{L}%
\alpha_{-2}^{L}|0\rangle.
\end{align}
The $s-t$ channel of these amplitudes can be calculated to be%
\begin{align}
A^{TTT}  &  =\int_{0}^{1}dx\cdot x^{k_{1}\cdot k_{2}}\left(  1-x\right)
^{k_{2}\cdot k_{3}}\cdot\left(  \frac{ie^{T}\cdot k_{1}}{x}-\frac{ie^{T}\cdot
k_{3}}{1-x}\right)  ^{3}\nonumber\\
&  \simeq-i\left(  \sqrt{-t}\right)  ^{3}\frac{\Gamma\left(  -\frac{s}%
{2}-1\right)  \Gamma\left(  -\frac{\tilde{t}}{2}-1\right)  }{\Gamma\left(
\frac{u}{2}+3\right)  }\cdot\left(  -\frac{1}{8}s^{3}+\frac{1}{2}s\right)  ,
\end{align}%
\begin{align}
A^{LTT}  &  =\int_{0}^{1}dx\cdot x^{k_{1}\cdot k_{2}}\left(  1-x\right)
^{k_{2}\cdot k_{3}}\cdot\left(  \frac{ie^{T}\cdot k_{1}}{x}-\frac{ie^{T}\cdot
k_{3}}{1-x}\right)  ^{2}\left(  \frac{ie^{L}\cdot k_{1}}{x}-\frac{ie^{L}\cdot
k_{3}}{1-x}\right) \nonumber\\
&  \simeq-i\left(  \sqrt{-t}\right)  ^{2}\left(  -\frac{1}{2M_{2}}\right)
\frac{\Gamma\left(  -\frac{s}{2}-1\right)  \Gamma\left(  -\frac{\tilde{t}}%
{2}-1\right)  }{\Gamma\left(  \frac{u}{2}+3\right)  }\cdot\left[  \frac{3}%
{4}s^{3}-\frac{t}{4}s^{2}-\left(  \frac{t}{2}+3\right)  s\right]  ,
\end{align}%
\begin{align}
A^{LLT}  &  =\int_{0}^{1}dx\cdot x^{k_{1}\cdot k_{2}}\left(  1-x\right)
^{k_{2}\cdot k_{3}}\cdot\left(  \frac{ie^{T}\cdot k_{1}}{x}-\frac{ie^{T}\cdot
k_{3}}{1-x}\right)  \left(  \frac{ie^{L}\cdot k_{1}}{x}-\frac{ie^{L}\cdot
k_{3}}{1-x}\right)  ^{2}\nonumber\\
&  \simeq-i\left(  \sqrt{-t}\right)  \left(  -\frac{1}{2M_{2}}\right)
^{2}\frac{\Gamma\left(  -\frac{s}{2}-1\right)  \Gamma\left(  -\frac{\tilde{t}%
}{2}-1\right)  }{\Gamma\left(  \frac{u}{2}+3\right)  }\nonumber\\
&  \cdot\left[  \left(  \frac{1}{4}t-\frac{9}{2}\right)  s^{3}+\left(
\frac{1}{4}t^{2}+\frac{7}{2}t\right)  s^{2}+\frac{\left(  t+6\right)  ^{2}}%
{2}s\right]  ,
\end{align}%
\begin{align}
A^{LLL}  &  =\int_{0}^{1}dx\cdot x^{k_{1}\cdot k_{2}}\left(  1-x\right)
^{k_{2}\cdot k_{3}}\cdot\left(  \frac{ie^{L}\cdot k_{1}}{x}-\frac{ie^{L}\cdot
k_{3}}{1-x}\right)  ^{3}\nonumber\\
&  \simeq-i\left(  -\frac{1}{2M_{2}}\right)  ^{3}\frac{\Gamma\left(  -\frac
{s}{2}-1\right)  \Gamma\left(  -\frac{\tilde{t}}{2}-1\right)  }{\Gamma\left(
\frac{u}{2}+3\right)  }\nonumber\\
&  \cdot\left[  -\left(  \frac{11}{2}t-27\right)  s^{3}-6\left(
t^{2}+6t\right)  s^{2}-\frac{\left(  t+6\right)  ^{3}}{2}s\right]  ,
\end{align}%
\begin{align}
A^{TT}  &  =\int_{0}^{1}dx\cdot x^{k_{1}\cdot k_{2}}\left(  1-x\right)
^{k_{2}\cdot k_{3}}\cdot\left(  \frac{ie^{T}\cdot k_{1}}{x}-\frac{ie^{T}\cdot
k_{3}}{1-x}\right)  \left[  \frac{e^{T}\cdot k_{1}}{x^{2}}+\frac{e^{T}\cdot
k_{3}}{\left(  1-x\right)  ^{2}}\right] \nonumber\\
&  \simeq-i\left(  \sqrt{-t}\right)  ^{2}\frac{\Gamma\left(  -\frac{s}%
{2}-1\right)  \Gamma\left(  -\frac{\tilde{t}}{2}-1\right)  }{\Gamma\left(
\frac{u}{2}+3\right)  }\left(  -\frac{1}{8}s^{3}+\frac{1}{2}s\right)  ,
\end{align}%
\begin{align}
A^{TL}  &  =\int_{0}^{1}dx\cdot x^{k_{1}\cdot k_{2}}\left(  1-x\right)
^{k_{2}\cdot k_{3}}\cdot\left(  \frac{ie^{T}\cdot k_{1}}{x}-\frac{ie^{T}\cdot
k_{3}}{1-x}\right)  \left[  \frac{e^{L}\cdot k_{1}}{x^{2}}+\frac{e^{L}\cdot
k_{3}}{\left(  1-x\right)  ^{2}}\right] \nonumber\\
&  \simeq i\left(  \sqrt{-t}\right)  \left(  -\frac{1}{2M_{2}}\right)
\frac{\Gamma\left(  -\frac{s}{2}-1\right)  \Gamma\left(  -\frac{\tilde{t}}%
{2}-1\right)  }{\Gamma\left(  \frac{u}{2}+3\right)  }\nonumber\\
&  \cdot\left[  -\left(  \frac{1}{8}t+\frac{3}{4}\right)  s^{3}-\frac{1}%
{8}\left(  t^{2}-2t\right)  s^{2}-\left(  \frac{1}{4}t^{2}-t-3\right)
s\right]  ,
\end{align}%
\begin{align}
A^{LT}  &  =\int_{0}^{1}dx\cdot x^{k_{1}\cdot k_{2}}\left(  1-x\right)
^{k_{2}\cdot k_{3}}\cdot\left(  \frac{ie^{L}\cdot k_{1}}{x}-\frac{ie^{L}\cdot
k_{3}}{1-x}\right)  \left[  \frac{e^{T}\cdot k_{1}}{x^{2}}+\frac{e^{T}\cdot
k_{3}}{\left(  1-x\right)  ^{2}}\right] \nonumber\\
&  \simeq i\left(  \sqrt{-t}\right)  \left(  -\frac{1}{2M_{2}}\right)
\frac{\Gamma\left(  -\frac{s}{2}-1\right)  \Gamma\left(  -\frac{\tilde{t}}%
{2}-1\right)  }{\Gamma\left(  \frac{u}{2}+3\right)  }\cdot\left[  \frac{3}%
{4}s^{3}-\frac{t}{4}s^{2}-\left(  \frac{t}{2}+3\right)  s\right]  ,
\end{align}
and%
\begin{align}
A^{LL}  &  =\int_{0}^{1}dx\cdot x^{k_{1}\cdot k_{2}}\left(  1-x\right)
^{k_{2}\cdot k_{3}}\cdot\left(  \frac{ie^{L}\cdot k_{1}}{x}-\frac{ie^{L}\cdot
k_{3}}{1-x}\right)  \left[  \frac{e^{L}\cdot k_{1}}{x^{2}}+\frac{e^{L}\cdot
k_{3}}{\left(  1-x\right)  ^{2}}\right] \nonumber\\
&  \simeq i\left(  -\frac{1}{2M_{2}}\right)  ^{2}\frac{\Gamma\left(  -\frac
{s}{2}-1\right)  \Gamma\left(  -\frac{\tilde{t}}{2}-1\right)  }{\Gamma\left(
\frac{u}{2}+3\right)  }\nonumber\\
&  \cdot\left[  \left(  \frac{3}{4}t+\frac{9}{2}\right)  s^{3}+\left(
t^{2}-4t\right)  s^{2}+\left(  \frac{1}{4}t^{3}+\frac{1}{2}t^{2}-9t-18\right)
s\right]  .
\end{align}
From the above calculation, one can easily see that all the amplitudes are in
the same leading order $\left(  \sim s^{3}\right)  $\ in the RR, while in the
GR only $A^{TTT}$, $A^{LLT}$ and $A^{TL}$ are in the leading order $\left(
\sim t^{3/2}s^{3}\text{ or }t^{5/2}s^{2}\right)  $, all other amplitudes are
in the subleading orders. On the other hand, one notes that, for example, the
term $\sim\sqrt{-t}t^{2}s^{2}$ in $A^{LLT}$ and $A^{TL}$ are in the leading
order in the GR, but are in the subleading order in the RR. On the contrary,
the terms $\sqrt{-t}s^{3}$ in $A^{LLT}$ and $A^{TL}$ are in the subleading
order in the GR, but are in the leading order in the RR. These observations
suggest that the high energy string scattering amplitudes in the GR and RR
contain information complementary to each other.

One can now see that the number of high energy scattering amplitudes in the RR
is much more numerous than that of GR. One important observation for high
energy amplitudes in the RR is for those amplitudes with the same structure as
those of the GR in Eq.(\ref{relevant states}). For these amplitudes, the
relative ratios of the coefficients of the highest power of $t$ in the leading
order amplitudes in the RR can be calculated to be
\begin{align}
A^{TTT}  &  =-i\left(  \sqrt{-t}\right)  \frac{\Gamma\left(  -\frac{s}%
{2}-1\right)  \Gamma\left(  -\frac{\tilde{t}}{2}-1\right)  }{\Gamma\left(
\frac{u}{2}+3\right)  }\cdot\left(  \frac{1}{8}ts^{3}\right)  \sim\frac{1}%
{8},\\
A^{LLT}  &  =-i\left(  \sqrt{-t}\right)  \left(  -\frac{1}{2M_{2}}\right)
^{2}\frac{\Gamma\left(  -\frac{s}{2}-1\right)  \Gamma\left(  -\frac{\tilde{t}%
}{2}-1\right)  }{\Gamma\left(  \frac{u}{2}+3\right)  }\left(  \frac{1}%
{4}ts^{3}\right)  \sim\frac{1}{64},\label{TLL}\\
A^{TL}  &  =i\left(  \sqrt{-t}\right)  \left(  -\frac{1}{2M_{2}}\right)
\frac{\Gamma\left(  -\frac{s}{2}-1\right)  \Gamma\left(  -\frac{\tilde{t}}%
{2}-1\right)  }{\Gamma\left(  \frac{u}{2}+3\right)  }\cdot\left(  -\frac{1}%
{8}ts^{3}\right)  \sim-\frac{1}{32}, \label{TL}%
\end{align}
which reproduces the ratios in the GR in Eq.(\ref{CL}). Note that the
symmetrized and anti-symmetrized amplitudes are defined as%
\begin{align}
T^{\left(  TL\right)  }  &  =\frac{1}{2}\left(  T^{TL}+T^{LT}\right)  ,\\
T^{\left[  TL\right]  }  &  =\frac{1}{2}\left(  T^{TL}-T^{LT}\right)  ;
\end{align}
and similarly for the amplitudes $A^{\left(  TL\right)  }$ and $A^{\left[
TL\right]  }$ in the RR. Note that $T^{LT}$ $\sim(\alpha_{-1}^{L})(\alpha
_{-2}^{T})|0\rangle$ in the GR is of subleading order in energy, while
$A^{LT}$ in the RR is of leading order in energy. However, the contribution of
the amplitude $A^{LT}$ to $A^{\left(  TL\right)  }$ and $A^{\left[  TL\right]
}$ in the RR will not affect the ratios calculated above. As we will see in
section IV, this interesting result can be generalized to all mass levels in
the string spectrum.%

\setcounter{equation}{0}
\renewcommand{\theequation}{\arabic{section}.\arabic{equation}}%

\section{General Mass Levels}

In this section, we calculate high energy string scattering amplitudes in the
RR for the arbitrary mass levels. Instead of states in
Eq.(\ref{relevant states}) for the GR, one can easily argue that the most
general string states one needs to consider at each fixed mass level
$N=\sum_{n,m}nk_{n}+mq_{m}$ for the RR are%
\begin{equation}
\left\vert k_{n},q_{m}\right\rangle =\prod_{n>0}(\alpha_{-n}^{T})^{k_{n}}%
\prod_{m>0}(\alpha_{-m}^{L})^{q_{m}}|0\rangle.
\end{equation}
It seems that both the saddle-point method and the method of decoupling of
high energy ZNS adopted in the calculation of GR do not apply to the case of
RR. However the calculation is still manageable, and the general formula for
the high energy scattering amplitudes in the RR can be written down
explicitly. In fact, by the simple kinematics $e^{T}\cdot k_{1}=0$, and the
energy power counting of the string amplitudes, we end up with the following
rules to simplify the calculation for the leading order amplitudes in the RR:
\begin{align}
&  \alpha_{-n}^{T}:\quad\text{1 term (contraction of $ik_{3}\cdot X$ with
$\varepsilon_{T}\cdot\partial^{n}X$),}\\
&  \alpha_{-n}^{L}:%
\begin{cases}
n>1,\quad\text{1 term}\\
n=1\quad\text{2 terms}\text{ (contraction of $ik_{1}\cdot X$ and $ik_{3}\cdot
X$ with $\varepsilon_{L}\cdot\partial^{n}X$).}%
\end{cases}
\end{align}
The $s-t$ channel scattering amplitudes of this state with three other
tachyonic states can be calculated to be
\begin{align}
A^{(k_{n},q_{m})}  &  =\int_{0}^{1}dx\,x^{k_{1}\cdot k_{2}}(1-x)^{k_{2}\cdot
k_{3}}\left[  \frac{ie^{L}\cdot k_{1}}{-x}+\frac{ie^{L}\cdot k_{3}}%
{1-x}\right]  ^{q_{1}}\nonumber\\
&  \cdot\prod_{n=1}\left[  \frac{ie^{T}\cdot k_{3}\,(n-1)!}{(1-x)^{n}}\right]
^{k_{n}}\prod_{m=2}\left[  \frac{ie^{L}\cdot k_{3}\,(m-1)!}{(1-x)^{m}}\right]
^{q_{m}}\nonumber\\
&  =\left(  \frac{-i\tilde{t}^{\prime}}{2M_{2}}\right)  ^{q_{1}}\sum
_{j=0}^{q_{1}}{\binom{q_{1}}{j}}\left(  \frac{s}{-\tilde{t}}\right)  ^{j}%
\int_{0}^{1}dxx^{k_{1}\cdot k_{2}-j}(1-x)^{k_{2}\cdot k_{3}+j-\sum
_{n,m}(nk_{n}+mq_{m})}\nonumber\\
&  \cdot\prod_{n=1}\left[  i\sqrt{-t}(n-1)!\right]  ^{k_{n}}\prod_{m=2}\left[
i\tilde{t}^{\prime}(m-1)!\left(  -\frac{1}{2M_{2}}\right)  \right]  ^{q_{m}%
}\nonumber\\
&  =\left(  \frac{-i\tilde{t}^{\prime}}{2M_{2}}\right)  ^{q_{1}}\sum
_{j=0}^{q_{1}}{\binom{q_{1}}{j}}\left(  \frac{s}{-\tilde{t}}\right)
^{j}B\left(  k_{1}\cdot k_{2}-j+1\,,\,k_{2}\cdot k_{3}+j-N+1\right)
\nonumber\\
&  \cdot\prod_{n=1}\left[  i\sqrt{-t}(n-1)!\right]  ^{k_{n}}\prod_{m=2}\left[
i\tilde{t}^{\prime}(m-1)!\left(  -\frac{1}{2M_{2}}\right)  \right]  ^{q_{m}}.
\end{align}
The Beta function above can be approximated in the large $s$, but fixed $t$
limit as follows
\begin{align}
&  B\left(  k_{1}\cdot k_{2}-j+1,k_{2}\cdot k_{3}+j-N+1\right) \nonumber\\
&  =B\left(  -1-\frac{s}{2}+N-j,-1-\frac{t}{2}+j\right) \nonumber\\
&  =\frac{\Gamma(-1-\frac{s}{2}+N-j)\Gamma(-1-\frac{t}{2}+j)}{\Gamma(\frac
{u}{2}+2)}\nonumber\\
&  \approx B\left(  -1-\frac{1}{2}s,-1-\frac{t}{2}\right)  \left(  -1-\frac
{s}{2}\right)  ^{N-j}\left(  \frac{u}{2}+2\right)  ^{-N}\left(  -1-\frac{t}%
{2}\right)  _{j}\nonumber\\
&  \approx B\left(  -1-\frac{1}{2}s,-1-\frac{t}{2}\right)  \left(  -\frac
{s}{2}\right)  ^{-j}\left(  -1-\frac{t}{2}\right)  _{j}.
\end{align}
where%
\begin{equation}
(a)_{j}=a(a+1)(a+2)...(a+j-1)
\end{equation}
is the Pochhammer symbol. The leading order amplitude in the RR can then be
written as%
\begin{align}
A^{(k_{n},q_{m})}  &  =\left(  \frac{-i\tilde{t}^{\prime}}{2M_{2}}\right)
^{q_{1}}B\left(  -1-\frac{1}{2}s,-1-\frac{t}{2}\right)  \sum_{j=0}^{q_{1}%
}{\binom{q_{1}}{j}}\left(  \frac{2}{\tilde{t}^{\prime}}\right)  ^{j}\left(
-1-\frac{t}{2}\right)  _{j}\nonumber\\
&  \cdot\prod_{n=1}\left[  i\sqrt{-t}(n-1)!\right]  ^{k_{n}}\prod_{m=2}\left[
i\tilde{t}^{\prime}(m-1)!\left(  -\frac{1}{2M_{2}}\right)  \right]  ^{q_{m}},
\label{A}%
\end{align}
which is UV power-law behaved as expected. The summation in eq. (\ref{A}) can
be represented by the Kummer function of the second kind $U$ as follows,
\begin{equation}
\sum_{j=0}^{p}{\binom{p}{j}}\left(  \frac{2}{\tilde{t}^{\prime}}\right)
^{j}\left(  -1-\frac{t}{2}\right)  _{j}=2^{p}(\tilde{t}^{\prime}%
)^{-p}\ U\left(  -p,\frac{t}{2}+2-p,\frac{\tilde{t}^{\prime}}{2}\right)  ...
\label{equality}%
\end{equation}
Finally, the amplitudes can be written as
\begin{align}
A^{(k_{n},q_{m})}  &  =\left(  -\frac{i}{M_{2}}\right)  ^{q_{1}}U\left(
-q_{1},\frac{t}{2}+2-q_{1},\frac{\tilde{t}^{\prime}}{2}\right)  B\left(
-1-\frac{s}{2},-1-\frac{t}{2}\right) \nonumber\\
&  \cdot\prod_{n=1}\left[  i\sqrt{-t}(n-1)!\right]  ^{k_{n}}\prod_{m=2}\left[
i\tilde{t}^{\prime}(m-1)!\left(  -\frac{1}{2M_{2}}\right)  \right]  ^{q_{m}}.
\label{general amplitude}%
\end{align}
In the above, $U$ is the Kummer function of the second kind and is defined to
be%
\begin{equation}
U(a,c,x)=\frac{\pi}{\sin\pi c}\left[  \frac{M(a,c,x)}{(a-c)!(c-1)!}%
-\frac{x^{1-c}M(a+1-c,2-c,x)}{(a-1)!(1-c)!}\right]  \text{ \ }(c\neq2,3,4...)
\end{equation}
where $M(a,c,x)=\sum_{j=0}^{\infty}\frac{(a)_{j}}{(c)_{j}}\frac{x^{j}}{j!}$ is
the Kummer function of the first kind. $U$ and $M$ are the two solutions of
the Kummer Equation%
\begin{equation}
xy^{^{\prime\prime}}(x)+(c-x)y^{\prime}(x)-ay(x)=0.
\end{equation}
It is crucial to note that $c=\frac{t}{2}+2-q_{1},$ and is not a constant as
in the usual case, so $U$ in Eq.(\ref{general amplitude}) is not a solution of
the Kummer equation. This will make our analysis in the next section more
complicated as we will see soon. On the contrary, since $a=-q_{1}$ an integer,
the Kummer function in Eq.(\ref{equality}) terminated to be a finite sum. This
will simplify the manipulation of Kummer function used in this paper.%

\setcounter{equation}{0}
\renewcommand{\theequation}{\arabic{section}.\arabic{equation}}%

\section{Reproducing the GR ratios in the RR}

In section II, we have learned that the relative coefficients of the highest
power $t$ terms in the leading order amplitudes in the RR can reproduce the
ratios of the amplitudes in the GR for the mass level $M_{2}^{2}=4$. Now we
are going to generalize the calculation to the string states of the arbitrary
mass levels. The leading order amplitudes of string states in the RR, which
share the same structure as Eq.(\ref{relevant states}) in the GR can be
written as%
\begin{align}
A^{(N,2m,q)} &  =B\left(  -1-\frac{s}{2},-1-\frac{t}{2}\right)  \sqrt
{-t}^{N-2m-2q}\left(  \frac{1}{2M_{2}}\right)  ^{2m+q}\nonumber\\
&  2^{2m}(\tilde{t}^{\prime})^{q}U\left(  -2m\,,\,\frac{t}{2}+2-2m\,,\,\frac
{\tilde{t}^{\prime}}{2}\right)  .\label{RAM}%
\end{align}
It is important to note that there is no linear relation among high energy
string scattering amplitudes of different string states for each fixed mass
level in the RR as can be seen from Eq.(\ref{RAM}). This is very different
from the result in the GR. In other words, the ratios $A^{(N,2m,q)}%
/A^{(N,0,0)}$ are $t$-dependent functions. As was done in section II for the
mass level $M_{2}^{2}=4$, we can extract the coefficients of the highest power
of $t$ in $A^{(N,2m,q)}/A^{(N,0,0)}$. We can use the identity of the Kummer
function in Eq.(\ref{equality}) to calculate
\begin{equation}
\frac{A^{(N,2m,q)}}{A^{(n,0,0)}}=(-1)^{q}\left(  \frac{1}{2M_{2}}\right)
^{2m+q}(-t)^{m}\sum_{j=0}^{2m}(-2m)_{j}\left(  -1-\frac{t}{2}\right)
_{j}\frac{(-2/t)^{j}}{j!}+\mathit{O}\left\{  \left(  \frac{1}{t}\right)
^{m+1}\right\}  .\label{13}%
\end{equation}
where we have replaced $\tilde{t}^{\prime}$ by $t$ as $t$ is large. If the
leading order coefficients in Eq.(\ref{13}) extracted from the high energy
string scattering amplitudes in the RR are to be identified with the ratios
calculated previously among high energy string scattering amplitudes in the GR
in Eq.(\ref{ratios}), we need the following identity
\begin{align}
&  \sum_{j=0}^{2m}(-2m)_{j}\left(  -1-\frac{t}{2}\right)  _{j}\frac
{(-2/t)^{j}}{j!}\nonumber\\
&  =0(-t)^{0}+0(-t)^{-1}+...+0(-t)^{-m+1}+\frac{(2m)!}{m!}(-t)^{-m}%
+\mathit{O}\left\{  \left(  \frac{1}{t}\right)  ^{m+1}\right\}  ..\label{ID}%
\end{align}
The coefficient of the term $\mathit{O}\left\{  \left(  1/t\right)
^{m+1}\right\}  $ in Eq.(\ref{ID}) is irrelevant for our discussion. The proof
of Eq.(\ref{ID}) suggested by string theory calculation turns out to be
nontrivial mathematically. Presumably, the difficulty of the rigorous proof of
Eq.(\ref{ID}) is associated with the unusual non-constant $c$ in the argument
of Kummer function in Eq.(\ref{RAM}) as mentioned above. We first rewrite the
summation on the left hand side of Eq.(\ref{ID}) as
\begin{align}
&  \sum_{j=0}^{2m}(-2m)_{j}\left(  -1-\frac{t}{2}\right)  \left(  -\frac{t}%
{2}\right)  _{j-1}\left(  -\frac{2}{t}\right)  ^{j}\frac{1}{j!}\nonumber\\
&  =\sum_{j=0}^{2m}(-2m)_{j}\left(  -1-\frac{t}{2}\right)  \sum_{k=0}%
^{j-1}(-1)^{j-1-k}s(j-1,k)\left(  -\frac{t}{2}\right)  ^{k}\left(  -\frac
{2}{t}\right)  ^{j}\frac{1}{j!}\nonumber\\
&  =\sum_{j=0}^{2m}(-2m)_{j}\sum_{k=0}^{j-1}(-1)^{j-k}s(j-1,k)(-1)^{j+k}%
2^{j-k}t^{k-j}\frac{1}{j!}\nonumber\\
&  +\frac{t}{2}\sum_{j=0}^{2m}(-2m)_{j}\sum_{k=0}^{j-1}(-1)^{j-k}%
s(j-1,k)(-1)^{j+k}2^{j-k}t^{k-j}\frac{1}{j!}.
\end{align}
In the above equation, we take $k=j-m$ for the first term and $k=j-m-1$ for
the second term. The equation then reduces to%
\begin{align}
&  \Longrightarrow\sum_{j=m}^{2m}(-2m)_{j}s(j-1,j-m)\frac{2^{m}}{j!}%
+\sum_{j=m+1}^{2m}(-2m)_{j}s(j-1,j-m-1)\frac{2^{m}}{j!}\nonumber\\
&  =2^{m}\sum_{j=0}^{m}(-1)^{j+m}{\binom{2m}{j+m}}s(j+m-1,j)\nonumber\\
&  +2^{m}\sum_{j=1}^{m}(-1)^{j+m}{\binom{2m}{j+m}}s(j+m-1,j-1)
\end{align}
where we have used the signed Stirling number of the first kind $s(n,k)$ to
expand the Pochhammer symbol. The definition of $s(n,k)$ is
\begin{equation}
(x)_{n}=\sum_{k=0}^{n}(-1)^{n-k}s(n,k)x^{k}.
\end{equation}
Thus the leading order nontrivial identity of Eq.(\ref{ID}) can be written as
($m\geqslant0$)
\begin{equation}
f(m)\equiv\sum_{j=0}^{m}(-1)^{j}{\binom{2m}{j+m}}\left[
s(j+m-1,j-1)+s(j+m-1,j)\right]  =(2m-1)!!\label{lead}%
\end{equation}
where we have used the convention that
\begin{equation}
s(m-1,-1)%
\begin{cases}
=0\quad,\text{ for $m\geqslant1$}\\
=1\quad,\text{ for $m=0$}%
\end{cases}
\quad,s(-1,0)=0.
\end{equation}
With the help of the algorithm developed by Mkauers in 2007 \cite{MK}, this
identity can be proved. The point is that we can find a recurrence relation of
$f(m)$ by his algorithm. However, to utilize the algorithm, we need to
introduce an auxiliary variable $u$ and define
\begin{align}
f(u,m) &  \equiv\sum_{j=0}^{m+u}(-1)^{j}{\binom{2m+u}{j+m}}\left[
s(j+m-1,j-1)+s(j+m-1,j)\right]  \nonumber\\
&  \equiv f_{1}(u,m)+f_{2}(u,m)
\end{align}
where $f_{1}$ and $f_{2}$ are the two summations, each with one Stirling
number, and $f(0,m)=f(m)$. By the algorithm, we can prove that both $f_{1}$,
$f_{2}$ satisfy the following recurrence relation \cite{MK}
\begin{equation}
-(1+2m+u)f(u,m)+(2m+u)f(u+1,m)+f(u,m+1)=0,\label{recurrence}%
\end{equation}
hence, so is $f.$ Eq.(\ref{recurrence}) is the most nontrivial step to prove
Eq.(\ref{lead}). Now, note that
\begin{equation}
f(u,0)=\sum_{j=0}^{u}(-1)^{j}{\binom{u}{j}}=%
\begin{cases}
1\quad,u=0\\
0\quad,u>0
\end{cases}
..
\end{equation}
Using the recurrence relation Eq.(\ref{recurrence}) and substituting
$(u,m)=(1,0),(2,0)\cdots$, one can prove that
\begin{equation}
f(u,1)=0,\quad\forall u>0.
\end{equation}
Similarly, by substituting $(u,m)=(1,1),(2,1),(3,1)\cdots$, one can get
$f(u,2)=0,\forall u>0$. In general, we have
\begin{equation}
f(u,m)=0,\quad\forall u>0.
\end{equation}
Finally we substitute $u=0$ in the Eq.(\ref{recurrence}) to obtain%
\begin{equation}
-(1+2m)f(0,m)+2mf(1,m)+f(0,m+1)=0,
\end{equation}
which implies%
\begin{equation}
f(m+1)=(2m+1)f(m).
\end{equation}
Eq.(\ref{lead}) is thus proved by mathematical induction.

The vanishing of the coefficients of $(-t)^{0},(-t)^{-1},...,(-t)^{-m+1}$
terms on the LHS of Eq.(\ref{ID}) means, for $1\leqslant i\leqslant m$,
\begin{align}
g(m,i)  &  \equiv\sum_{j=0}^{m+i}(-1)^{j-i}{\binom{2m}{j+m-i}}\left[
s(j+m-1-i,j)+s(j+m-1-i,j-1)\right] \nonumber\\
&  =0. \label{19}%
\end{align}
To prove this identity, we need the recurrence relation \cite{MK}
\begin{align}
-  &  2(1+m)^{2}(1+2m)g(m,i)+(2+7m+4m^{2})g(m+1,i)\nonumber\\
-  &  2m(1+m)(1+2m)g(m+1,i+1)-mg(m+2,i)=0. \label{20}%
\end{align}
Putting $i=0,1,2..$, and using the fact we have just proved, i.e.
$g(m+1,0)=(2m+1)g(m,0)$, one can show that
\begin{equation}
g(m,i)=0\quad\text{for $1\leqslant{i}\leqslant{m.}$} \label{21}%
\end{equation}
Eq.(\ref{ID}) is finally proved. We thus have shown that for those leading
order high energy amplitudes $A^{(N,2m,q)}$ in the RR with the same type of
$(N,2m,q)$ as those of GR, we can extract from them the ratios $T^{(N,2m,q)}%
/T^{(N,0,0)}$ in the GR by using the Kummer function. Mathematically, the
proof of this result turns out to be highly nontrivial and is based on a
summation algorithm for Stirling number identity derived by Mkauers \cite{MK}.
It is very interesting to see that the identity in Eq.(\ref{ID}) suggested by
string scattering amplitude calculation can be rigorously proved by a totally
different but sophisticated mathematical method. In the next section, we
discuss the generalization to subleading order amplitudes in the RR.%

\setcounter{equation}{0}
\renewcommand{\theequation}{\arabic{section}.\arabic{equation}}%

\section{Subleading Orders}

In this section, we calculate the next few subleading order amplitudes in the
RR for the mass level $M_{2}^{2}=4,6$. We will see that the ratios in
Eqs.(\ref{CL}) and (\ref{CL2}) persist to subleading order amplitudes in the
RR. For the even mass levels with $(N-1)=\frac{M_{2}^{2}}{2}$= even, we
conjecture and give evidences that the existence of these ratios in the RR
persists to all orders in the Regge expansion of all high energy string
scattering amplitudes . For the odd mass levels with $(N-1)=\frac{M_{2}^{2}%
}{2}$= odd, the existence of these ratios will show up only in the first
[N/2]+1 terms in the Regge expansion of the amplitudes.

We will extend the kinematic relations in the RR to the subleading orders. We
first express all kinematic variables in terms of $s$ and $t$, and then expand
all relevant quantities in $s:$
\begin{align}
E_{1}  &  =\frac{s-(m_{2}^{2}+2)}{2\sqrt{2}},\\
E_{2}  &  =\frac{s+(m_{2}^{2}+2)}{2\sqrt{2}},\\
|\mathbf{k_{2}}|  &  =\sqrt{E_{1}^{2}+2},\quad|\mathbf{K_{3}}|=\sqrt{\frac
{s}{4}+2};
\end{align}%
\begin{equation}
e_{P}\cdot k_{1}=-\frac{1}{2m_{2}}s+\left(  -\frac{1}{m_{2}}+\frac{m_{2}}%
{2}\right)  ,\quad(\text{exact})
\end{equation}%
\begin{align}
e_{L}\cdot k_{1}  &  =-\frac{1}{2m_{2}}s+\left(  -\frac{1}{m_{2}}+\frac{m_{2}%
}{2}\right)  -{2m_{2}}s^{-1}-2m_{2}(m_{2}^{2}-2)s^{-2}\nonumber\\
&  -2m_{2}(m_{2}^{4}-6m_{2}^{2}+4)s^{-3}-2m_{2}(m_{2}^{6}-12m_{2}^{4}%
+24m_{2}^{2}-8)s^{-4}+O(s^{-5}),
\end{align}%
\begin{equation}
e_{T}\cdot k_{1}=0.
\end{equation}
A key step is to express the scattering angle $\theta$ in terms of $s$ and
$t$. This can be achieved by solving
\begin{equation}
t=-\left(  -(E_{2}-\frac{\sqrt{s}}{2})^{2}+(|\mathbf{k_{2}}|-|\mathbf{k_{3}%
}|\cos\theta)^{2}+|\mathbf{k}_{3}|^{2}\sin^{2}\theta\right)
\end{equation}
to obtain%
\begin{equation}
\theta=\arccos\left(  \frac{s+2t-m_{2}^{2}+6}{\sqrt{s+8}\sqrt{\frac
{(s+2)^{2}-2(s-2)m_{2}^{2}+m_{2}^{4}}{s}}}\right)  .\text{ (exact)}
\label{t-angle}%
\end{equation}
One can then calculate the following expansions
\begin{equation}
e_{P}\cdot k_{3}=\frac{1}{m_{2}}(E_{2}\frac{\sqrt{s}}{2}-|\mathbf{k_{2}%
}||\mathbf{k_{3}}|\cos\theta)=-\frac{t+2-m_{2}^{2}}{2m_{2}},
\end{equation}%
\begin{align}
e_{L}\cdot k_{3}  &  =\frac{1}{m_{2}}(k_{2}\frac{\sqrt{2}}{2}-E_{2}k_{3}%
\cos\theta)\nonumber\\
&  =-\frac{t+2+m_{2}^{2}}{2m_{2}}-m_{2}ts^{-1}-m_{2}[-4(t+1)+m_{2}%
^{2}(t-2)]s^{-2}\nonumber\\
&  -m_{2}[4(4+3t)-12tm_{2}^{2}+(t-4)m_{2}^{4}]s^{-3}\nonumber\\
&  \quad-m_{2}[-16(3+2t)+24(2+3t)m_{2}^{2}\nonumber\\
&  -24(-1+t)m_{2}^{4}+(-6+t)m_{2}^{6}]s^{-4}+O(s^{-5}), \label{tpower1}%
\end{align}%
\begin{align}
e_{T}\cdot k_{3}  &  =-|\mathbf{k_{3}}|\sin\theta\nonumber\\
&  =-\sqrt{-t}-\frac{1}{2}\sqrt{-t}(2+t+m_{2}^{2})s^{-1}\nonumber\\
&  \quad-\frac{1}{8\sqrt{-t}}[32+52t+20t^{2}+t^{3}+(32+20t-6t^{2})m_{2}%
^{2}+(8-3t)m_{2}^{4}]s^{-2}\nonumber\\
&  \quad+\frac{1}{16\sqrt{-t}}[320+456t+188t^{2}+22t^{3}+t^{4}%
-(-224+36t+132t^{2}+5t^{3})m_{2}^{2}.\nonumber\\
&  \quad\quad\quad\quad\quad\quad+(-16-122t+15t^{2})m_{2}^{4}+(-24+5t)m_{2}%
^{6}]s^{-3}\nonumber\\
&  \quad+\frac{1}{128(-t)^{3/2}}[1024+12032t+16080t^{2}+7520t^{3}%
+1432t^{4}+136t^{5}+5t^{6}\nonumber\\
&  -4(-512-896t+2232t^{2}+1844t^{3}+170t^{4}+7t^{5})m_{2}^{2}\nonumber\\
&  +2(768-2240t-2372t^{2}+1172t^{3}+35t^{4})m_{2}^{4}\nonumber\\
&  -4(-128+288t-450t^{2}+35t^{3})m_{2}^{6}+(64+240t-35t^{2})m_{2}^{8}%
]s^{-4}+O(s^{-5}). \label{tpower2}%
\end{align}
We are now ready to calculate the expansions of the four amplitudes
$A_{TTT},A_{LLT},A_{[LT]},A_{(LT)}$ for the mass level $M_{2}^{2}=4$ to
subleading orders in $s$ in the RR. These are
\begin{equation}
A^{TTT}\sim\frac{1}{8}\sqrt{-t}ts^{3}+\frac{3}{16}\sqrt{-t}t(t+6)s^{2}%
+\frac{3t^{3}+84t^{2}-68t-864}{64}\sqrt{-t}\,s+O(1),
\end{equation}%
\begin{align}
A^{LLT}  &  \sim\frac{1}{64}\sqrt{-t}(t-6)s^{3}+\frac{3}{128}\sqrt{-t}%
(t^{2}-20t-12)s^{2}\nonumber\\
&  \quad\quad+\frac{3t^{3}-342t^{2}-92t+5016+1728(-t)^{-1/2}}{512}\sqrt
{-t}\,s+O(1),
\end{align}%
\begin{align}
A^{[LT]}  &  \sim-\frac{1}{64}\sqrt{-t}(t+2)s^{3}-\frac{3}{128}\sqrt
{-t}(t+2)^{2}s^{2}+O(s)\nonumber\\
&  \quad\quad-\frac{(3t-8)(t+6)^{2}[1-2(-t)^{-1/2}]}{512}\sqrt{-t}\,s+O(1),
\end{align}%
\begin{align}
A^{(LT)}  &  \sim-\frac{1}{64}\sqrt{-t}(t+10)s^{3}-\frac{1}{128}\sqrt
{-t}(3t^{2}+52t+60)s^{2}+O(s)\nonumber\\
&  \quad\quad-\frac{3[t^{3}+30t^{2}+76t-1080-960(-t)^{-1/2}]}{512}\sqrt
{-t}\,s+O(1).
\end{align}
One can now easily see that the ratios of the coefficients of the highest
power of $t$ in the leading order coefficient functions $\frac{1}{8}:\frac
{1}{64}:-\frac{1}{64}:-\frac{1}{64}$ agree with the ratios in the GR
$8:1:-1:-1$ calculated in Eq.(\ref{CL}) as expected. Moreover, one further
obeservation is that these ratios remain the same for the coefficients of the
highest power of $t$ in the subleading orders $(s^{2})$ $\frac{3}{16}:\frac
{3}{128}:-\frac{3}{128}:-\frac{3}{128}$ and $(s)$ $\frac{3}{64}:\frac{3}%
{512}:-\frac{3}{512}:-\frac{3}{512}$. We conjecture that these ratios persist
to all energy orders in the Regge expansion of the amplitudes. This is
consistent with the results of GR by taking both $s,-t\rightarrow\infty.$ For
the mass level $M_{2}^{2}=6$ \cite{ChanLee2}, the amplitudes can be calculated
to be
\begin{align}
A^{TTTT}  &  \sim\left(  \frac{s^{2}}{4}-s\right)  \left(  \frac{s^{2}}%
{4}-1\right)  (e^{T}\cdot k_{3})^{4}\nonumber\\
&  =\frac{t^{2}}{16}s^{4}+\frac{t^{2}(t+6)}{8}s^{3}+\frac{t(t^{3}%
+24t^{2}-4t-256)}{16}s^{2}\nonumber\\
&  +\frac{t(3t^{3}-2t^{2}-396t-768)}{4}s-\left(  \frac{t^{4}}{4}%
+166t^{3}+960t^{2}-64t-1024\right)  s^{0}\nonumber\\
&  +(-83t^{4}-1536t^{3}+384t^{2}+21248t+12288)s^{-1}+O(s^{-2}),
\end{align}%
\begin{align}
A^{TTLL}  &  \sim\left(  \frac{s^{2}}{4}-s\right)  \left(  \frac{s^{2}}%
{4}-1\right)  (e^{T}\cdot k_{3})^{2}(e^{L}\cdot{k}_{3})^{2}\nonumber\\
&  +\frac{3st}{2}\left(  \frac{s}{2}+1\right)  \left(  \frac{t}{2}+1\right)
(e^{L}\cdot k_{1})^{2}(e^{T}\cdot k_{3})^{2}\times\frac{1}{6}\nonumber\\
&  -s\left(  \frac{s^{2}}{4}-1\right)  (t+2)(e^{L}\cdot k_{1})(e^{L}\cdot
k_{3})(e^{T}\cdot k_{3})^{2}\times\frac{1}{2}\nonumber\\
&  =\frac{t(t-16)}{192}s^{4}+\frac{t(t^{2}-41t-32)}{96}s^{3}+\frac
{t^{4}-132t^{3}-328t^{2}+1984t+2048}{192}s^{2}\nonumber\\
&  +\left(  -\frac{11t^{4}}{32}-\frac{11t^{3}}{4}+\frac{163t^{2}}%
{3}+184t+\frac{128}{3}\right)  s^{1}\nonumber\\
&  +\left(  -\frac{11}{8}t^{4}+88t^{3}+744t^{2}+304t-1408\right)
s^{0}\nonumber\\
&  +4\left(  11t^{4}+280t^{3}+204t^{2}-4448t-4480\right)  s^{-1}+O(s^{-2}),
\end{align}%
\begin{align}
A^{LLLL}  &  \sim\left(  \frac{s^{2}}{4}-s\right)  \left(  \frac{s^{2}}%
{4}-1\right)  (e^{L}\cdot k_{3})^{4}-t\left(  \frac{t^{2}}{4}-1\right)
(s+2)(e^{L}\cdot k_{1})^{3}(e^{L}\cdot k_{3})\nonumber\\
&  +\frac{3st}{2}\left(  \frac{s}{2}+1\right)  \left(  \frac{t}{2}+1\right)
(e^{L}\cdot k_{1})^{2}(e^{L}\cdot k_{3})^{2}\nonumber\\
&  -s\left(  \frac{s^{2}}{4}-1\right)  (t+2)(e^{L}\cdot k_{1})(e^{L}\cdot
k_{3})^{3}\nonumber\\
&  +\left(  \frac{t^{2}}{4}-t\right)  \left(  \frac{t^{2}}{4}-1\right)
(e^{L}\cdot k_{1})^{4}\nonumber\\
&  =\frac{t(t-52)}{768}s^{4}+\frac{t(t^{2}-140t+256)}{384}s^{3}+\frac
{t^{4}-456t^{3}+2816t^{2}-512t-16384}{768}s^{2}\nonumber\\
&  \left(  -\frac{19t^{4}}{64}+6t^{3}-\frac{17t^{2}}{3}-176t-\frac{256}%
{3}\right)  s^{1}\nonumber\\
&  +(3t^{4}-10t^{3}-528t^{2}-672t+1792)s^{0}+O(s^{-1}),
\end{align}%
\begin{align}
A^{TTL}  &  \sim-\left(  \frac{s^{2}}{4}-s\right)  \left(  \frac{s^{2}}%
{4}-1\right)  (e^{T}\cdot k_{3})^{2}(e^{L}\cdot k_{3})\nonumber\\
&  -\frac{st}{4}\left(  \frac{s}{2}+1\right)  \left(  \frac{t}{2}+1\right)
(e^{T}\cdot k_{3})^{2}(e^{L}\cdot k_{1})\times\frac{1}{3}\nonumber\\
&  +s\left(  \frac{s^{2}}{4}-1\right)  \left(  \frac{t}{2}+1\right)
(e^{T}\cdot k_{3})^{2}(e^{L}\cdot k_{1})\times\frac{1}{3}\nonumber\\
&  =-\frac{(t+20)t}{96\sqrt{6}}s^{4}-\frac{t(t^{2}+31t+40)}{48\sqrt{6}}%
s^{3}-\frac{t^{4}+38t^{3}+224t^{2}-1520t-2560}{96\sqrt{6}}s^{2}\nonumber\\
&  +\frac{-3t^{4}-72t^{3}+2248t^{2}+12000t+5120}{48\sqrt{6}}s^{1}\nonumber\\
&  +\frac{67t^{3}+1194t^{2}+1344t-3712}{2\sqrt{6}}s^{0}+O(s^{-1}),
\end{align}%
\begin{align}
A^{LLL}  &  \sim-\left(  \frac{s^{2}}{4}-s\right)  \left(  \frac{s^{2}}%
{4}-1\right)  (e^{L}\cdot k_{3})^{3}+t\left(  \frac{t^{2}}{4}-1\right)
\left(  \frac{s}{2}+1\right)  (e^{L}\cdot k_{1})^{2}(e^{L}\cdot k_{3}%
)\nonumber\\
&  -\frac{st}{4}\left(  \frac{s}{2}+1\right)  \left(  \frac{t}{2}+1\right)
\left[  (e^{L}\cdot k_{1})^{2}(e^{L}\cdot k_{3})+(e^{L}\cdot k_{3})^{2}%
(e^{L}\cdot k_{1})\right] \nonumber\\
&  +s\left(  \frac{s^{2}}{4}-1\right)  \left(  \frac{t}{2}+1\right)
(e^{L}\cdot k_{3})^{2}(e^{L}\cdot k_{1})-\left(  \frac{t^{2}}{4}-t\right)
\left(  \frac{t^{2}}{4}-1\right)  (e^{L}\cdot k_{1})^{3}\nonumber\\
&  =-\frac{t^{2}-8t-128}{384\sqrt{6}}s^{4}-\frac{t^{3}-52t^{2}-412t+256}%
{192\sqrt{6}}s^{3}\nonumber\\
&  -\frac{t^{4}-236t^{3}-1272t^{2}+4832t+15872}{384\sqrt{6}}s^{2}\nonumber\\
&  +\frac{35t^{4}+50t^{3}-3008t^{2}-23728t-14848}{96\sqrt{6}}s^{1}\nonumber\\
&  -\frac{47t^{4}+1432t^{3}+24796t^{2}+40640t-101376}{48\sqrt{6}}%
s^{0}+O(s^{-1}),
\end{align}%
\begin{align}
\tilde{A}^{LT,T}  &  \sim-\left(  \frac{s^{2}}{4}-s\right)  \left(
\frac{s^{2}}{4}-1\right)  (e^{T}\cdot k_{3})^{2}(e^{L}\cdot k_{3}%
)\times0\nonumber\\
&  -\frac{st}{4}\left(  \frac{s}{2}+1\right)  \left(  \frac{t}{2}+1\right)
(e^{L}\cdot k_{1})(e^{T}\cdot k_{3})^{2}\times\frac{1}{2}\nonumber\\
&  +s\left(  \frac{s^{2}}{4}-1\right)  \left(  \frac{t}{2}+1\right)
(e^{T}\cdot k_{3})^{2}(e^{L}\cdot k_{1})\times\left(  -\frac{1}{4}\right)
\nonumber\\
&  =-\frac{t(t+2)}{64\sqrt{6}}s^{4}-\frac{t(t+2)^{2}}{32\sqrt{6}}s^{3}%
-\frac{t^{4}+12t^{3}+8t^{2}-152t-256}{64\sqrt{6}}s^{2}\nonumber\\
&  +\frac{-3t^{4}+196t^{2}+624t+512}{32\sqrt{6}}s^{1}+\sqrt{\frac{3}{8}%
}(5t^{3}+30t^{2}+24t-32)s^{0}+O(s^{-1}),
\end{align}%
\begin{align}
A^{LL}  &  \sim\left(  \frac{s^{2}}{4}-s\right)  \left(  \frac{s^{2}}%
{4}-1\right)  (e^{L}\cdot k_{3})^{2}+\frac{st}{2}\left(  \frac{s}{2}+1\right)
\left(  \frac{t}{2}+1\right)  (e^{L}\cdot k_{1})(e^{L}\cdot k_{3})\nonumber\\
&  +\left(  \frac{t^{2}}{4}-t\right)  \left(  \frac{t^{2}}{4}-1\right)
(e^{L}\cdot k_{1})^{2}\nonumber\\
&  =\frac{(t+8)^{2}}{384}s^{4}+\frac{(t^{3}+20t^{2}+80t-128)}{192}s^{3}%
+\frac{t^{4}+16t^{3}+96t^{2}-880t-3328}{384}s^{2}\nonumber\\
&  +\frac{-t^{4}+8t^{3}-110t^{2}-1648t-1408}{48}s^{1}\nonumber\\
&  +\frac{t^{4}-4t^{3}-202t^{2}-704t+1728}{6}s^{0}+O(s^{-1}).
\end{align}
In the above calculations, as in the case of $M_{2}^{2}=4$, we have ingored a
common overall factor which will be discussed in section VI. Note that the
ratios of the coefficients in the leading order $t$ for the energy orders
$s^{4},s^{3},s^{2}$ reproduced the GR ratios in Eq.(\ref{CL2}). However, the
subleading terms for orders $s^{1},s^{0}$ contain no GR ratios.
Mathematically, this is because the highest power of $t$ in the coefficient
functions of $s^{1}$ is $4$ rather than $5$, and those of $s^{0}$ is $4$
rather than $6$. This is because the power of $t$ in the kinematic relation
Eq.(\ref{tpower2}) can be as high as one wants if one goes to subleading
orders, while that of Eq.(\ref{tpower1}) is not. The $\sin\theta$ factor in
Eq.(\ref{tpower2}) contributes terms of higher order powers of $t$, while
$\cos\theta$ factor in in Eq.(\ref{tpower1}) does not. This can be seen from
the kinematic relation in Eq.(\ref{t-angle}). In general, one can easily show
that the $\sin\theta$ factor will contribute only for the even mass levels
with $(N-1)=\frac{M_{2}^{2}}{2}$= even. We thus conjecture that the existence
of the GR ratios in the RR persists to all orders in the Regge expansion of
all string amplitudes for the even mass level. For the odd mass levels with
$(N-1)=\frac{M_{2}^{2}}{2}$= odd, the existence of the GR ratios will show up
only in the first $[N/2]+1$ terms in the Regge expansion of the amplitudes. An
interesting question is whether this phenomena persists for the case of
superstring where GSO projection needs to be imposed.%

\setcounter{equation}{0}
\renewcommand{\theequation}{\arabic{section}.\arabic{equation}}%

\section{Universal Power Law Behavior}

In the discussion of section V, we ignored an overall common factor$\frac
{\Gamma(-1-s/2)\Gamma(-1-t/2)}{\Gamma(u/2+2)}$ of the amplitudes for mass
levels $M_{2}^{2}=4,6$. We paid attention only to the ratios among scattering
amplitudes of different string states. In this section, we calculate the high
energy behavior of string scattering amplitudes for string states at arbitrary
mass levels in the RR. The power law behavior $\sim s^{\alpha(t)}$ of the
four-tachyon amplitude in the RR is well known in the literature. Here we want
to generalize this result to string states at arbitrary mass levels. We can
use the saddle point method to calculate the leading term of gamma functions
in the RR
\begin{equation}
\frac{\Gamma(-1-s/2)\Gamma(-1-t/2)}{\Gamma(u/2+2)}=\frac{\Gamma(-1-s/2)\Gamma
(-1-t/2)}{\Gamma(-s/2-t/2+N-2)}\sim s^{t/2-N+1}\text{ \ (in the RR)}.
\end{equation}
Thus, the overall $s$-dependence in the amplitudes is of the form%
\begin{equation}
A^{(k_{n},q_{m})}\sim s^{\alpha(t)}\text{ \ \ (in the RR)} \label{universal}%
\end{equation}
where
\begin{equation}
\alpha(t)=\alpha(0)+\alpha^{\prime}t\text{, \ }\alpha(0)=1\text{ and }%
\alpha^{\prime}=1/2.
\end{equation}
This generalizes the high energy behavior of the four-tachyon amplitude in the
RR to string states at arbitrary mass levels. The new result here is that the
behavior is universal and is mass level independent. In fact, as a simple
application, one can also derive Eq.(\ref{universal}) directly from
Eq.(\ref{general amplitude}) by using%
\begin{equation}
B\left(  -1-\frac{s}{2},-1-\frac{t}{2}\right)  \sim s^{\alpha(t)}.\text{ \ (in
the RR)}%
\end{equation}
We conclude that the well known $\sim s^{\alpha(t)}$ power-law behavior of the
four tachyon string scattering amplitude in the RR can be extended to high
energy string scattering amplitudes of arbitrary string states.

%

\setcounter{equation}{0}
\renewcommand{\theequation}{\arabic{section}.\arabic{equation}}%

\section{Conclusion}

In this paper, we calculate high energy massive string scattering amplitudes
of 26D open bosonic string in the Regge regime (RR). It turns out that both
the saddle-point method and the method of decoupling of high energy ZNS
adopted in the calculation of GR \cite{ChanLee1,ChanLee2,
CHL,CHLTY,PRL,susy,Closed,Decay,Compact} do not apply to the case of RR.
However, the general formula for the high energy scattering amplitudes for
each fixed mass level in the RR can still be written down explicitly. We have
found that the number of high energy amplitudes for each fixed mass level in
the RR is much more numerous than that of Gross regime (GR) calculated
previously \cite{ChanLee1,ChanLee2, CHL,CHLTY,PRL,susy,Closed,Decay,Compact}.

On the other hand, there is no linear relation among scatterings in the RR in
contrast to the case of scatterings in the GR. Moreover, we discover that the
leading order amplitudes in the RR can be expressed in terms of the Kummer
function of the second kind. In particular, based on a summation algorithm for
Stirling number identity in the combinatoric number theory, we discover that
the ratios calculated previously among scattering amplitudes in the GR can be
extracted from this Kummer function in the RR. We conjecture and give
evidences that the existence of the GR ratios in the RR persists to all orders
in the Regge expansion of all string amplitudes for the even mass level with
$(N-1)=\frac{M_{2}^{2}}{2}$= even. For the odd mass levels with $(N-1)=\frac
{M_{2}^{2}}{2}$= odd, the existence of the GR ratios shows up only in the
first $[N/2]+1$ terms in the Regge expansion of the amplitudes. An interesting
question is whether this phenomena persists for the case of superstring where
GSO projection needs to be imposed. Finally, we demonstrate the universal
power-law behavior for all massive string scattering amplitudes in the RR.
This result generalizes the well known result for the case of high energy
four-point tachyon amplitudes.

\begin{acknowledgments}
This work is supported in part by the National Science Council, 50 billions
project of MOE and National Center for Theoretical Science, Taiwan. We
appreciated the correspondence of Dr. Manuel Mkauers at RISC, Austria for his
kind help of providing us with the rigorous proof of Eq.(\ref{lead}), and for
informing us reference \cite{MK}.
\end{acknowledgments}

\appendix%

\setcounter{equation}{0}
\renewcommand{\theequation}{\thesection.\arabic{equation}}%

\section{Kinematic Variables and Notations}

In this appendix, we list the expressions of the kinematic variables we used
in the evaluation of 4-point functions in this paper. For convenience, we take
the center of momentum frame and choose the momenta of particles 1 and 2 to be
along the $X^{1}$-direction. The high energy scattering plane is defined to be
on the $X^{1}-X^{2}$ plane.

\begin{figure}[ph]
\label{scattering} \setlength{\unitlength}{3pt}
\par
\begin{center}
\begin{picture}(100,100)(-50,-50)
{\large
\put(45,0){\vector(-1,0){42}} \put(-45,0){\vector(1,0){42}}
\put(2,2){\vector(1,1){30}} \put(-2,-2){\vector(-1,-1){30}}
\put(25,2){$k_1$} \put(-27,2){$k_2$} \put(11,20){$-k_3$}
\put(-24,-15){$-k_4$}
\put(40,0){\vector(0,-1){10}} \put(-40,0){\vector(0,1){10}}
\put(26,26){\vector(-1,1){7}} \put(-26,-26){\vector(1,-1){7}}
\put(36,-16){$e^{T}(1)$} \put(-44,15){$e^{T}(2)$}
\put(15,36){$e^{T}(3)$} \put(-18,-35){$e^{T}(4)$}
\qbezier(10,0)(10,4)(6,6) \put(12,4){$\theta$}
\put(-55,-45){Fig.1 Kinematic variables in the center of mass frame} }
\end{picture}
\end{center}
\end{figure}

The momenta of the four particles are%

\begin{align}
k_{1}  &  =\left(  +\sqrt{p^{2}+M_{1}^{2}},-p,0\right)  ,\\
k_{2}  &  =\left(  +\sqrt{p^{2}+M_{2}^{2}},+p,0\right)  ,\\
k_{3}  &  =\left(  -\sqrt{q^{2}+M_{3}^{2}},-q\cos\phi,-q\sin\theta\right)  ,\\
k_{4}  &  =\left(  -\sqrt{q^{2}+M_{4}^{2}},+q\cos\phi,+q\sin\theta\right)
\end{align}

where $p\equiv\left\vert \mathrm{\vec{p}}\right\vert $, $q\equiv\left\vert
\mathrm{\vec{q}}\right\vert $ and $k_{i}^{2}=-M_{i}^{2}$. In the calculation
of the string scattering amplitudes, we use the following formulas%

\begin{align}
-k_{1}\cdot k_{2}  &  =\sqrt{p^{2}+M_{1}^{2}}\cdot\sqrt{p^{2}+M_{2}^{2}}%
+p^{2}=\dfrac{1}{2}\left(  s-M_{1}^{2}-M_{2}^{2}\right)  ,\\
-k_{2}\cdot k_{3}  &  =-\sqrt{p^{2}+M_{2}^{2}}\cdot\sqrt{q^{2}+M_{3}^{2}%
}+pq\cos\theta=\dfrac{1}{2}\left(  t-M_{2}^{2}-M_{3}^{2}\right)  ,\\
-k_{1}\cdot k_{3}  &  =-\sqrt{p^{2}+M_{1}^{2}}\cdot\sqrt{q^{2}+M_{3}^{2}%
}-pq\cos\theta=\dfrac{1}{2}\left(  u-M_{1}^{2}-M_{3}^{2}\right)
\end{align}
where the Mandelstam variables are defined as usual with%

\begin{equation}
s+t+u=\sum_{i}M_{i}^{2}=2\left(  N-4\right)  .
\end{equation}
The center of mass energy $E$ is defined as%
\begin{equation}
E=\dfrac{1}{2}\left(  \sqrt{p^{2}+M_{1}^{2}}+\sqrt{p^{2}+M_{2}^{2}}\right)
=\dfrac{1}{2}\left(  \sqrt{q^{2}+M_{3}^{2}}+\sqrt{q^{2}+M_{4}^{2}}\right)  .
\end{equation}
We define the polarizations of the string state on the scattering plane as%

\begin{align}
e^{P}  &  =\frac{1}{M_{2}}\left(  \sqrt{p^{2}+M_{2}^{2}},p,0\right)  ,\\
e^{L}  &  =\frac{1}{M_{2}}\left(  p,\sqrt{p^{2}+M_{2}^{2}},0\right)  ,\\
e^{T}  &  =\left(  0,0,1\right)  .
\end{align}
The projections of the momenta on the scattering plane can be calculated as
(here we only list the ones we need for our calculations)%

\begin{align}
e^{P}\cdot k_{1}  &  =-\frac{1}{M_{2}}\left(  \sqrt{p^{2}+M_{1}^{2}}%
\sqrt{p^{2}+M_{2}^{2}}+p^{2}\right)  ,\label{A13}\\
e^{L}\cdot k_{1}  &  =-\frac{p}{M_{2}}\left(  \sqrt{p^{2}+M_{1}^{2}}%
+\sqrt{p^{2}+M_{2}^{2}}\right)  ,\\
e^{T}\cdot k_{1}  &  =0
\end{align}
and%
\begin{align}
e^{P}\cdot k_{3}  &  =\frac{1}{M_{2}}\left(  \sqrt{q^{2}+M_{3}^{2}}\sqrt
{p^{2}+M_{2}^{2}}-pq\cos\theta\right)  ,\\
e^{L}\cdot k_{3}  &  =\frac{1}{M_{2}}\left(  p\sqrt{q^{2}+M_{3}^{2}}%
-q\sqrt{p^{2}+M_{2}^{2}}\cos\theta\right)  ,\\
e^{T}\cdot k_{3}  &  =-q\sin\theta=-\sqrt{-t}. \label{A18}%
\end{align}

\end{document}